\documentclass[a4paper,11pt]{article}
\pdfoutput=1 

\usepackage{jcappub} 

\usepackage[T1]{fontenc} 

\usepackage{makeidx}
\usepackage{multirow}
\usepackage{multicol}
\usepackage[dvipsnames,svgnames,table]{xcolor}
\usepackage{graphics}
\usepackage{epstopdf}
\usepackage{ulem}
\usepackage{hyperref}
\usepackage{amsmath}
\usepackage[compat=1.0.0]{tikz-feynman}
\usepackage{amssymb}
\usepackage{float}
\usepackage{fancyhdr} 
\usepackage{enumitem}
\usepackage{tcolorbox}
\usepackage{lipsum}
\usepackage{bold-extra}
\usepackage{wrapfig}
\usepackage{caption}
\usepackage[numbers]{natbib}
\allowdisplaybreaks
\usepackage{upgreek}
\usepackage{color}

\usepackage{bm}					
\usepackage{booktabs} 

\newcommand{\be}{\begin{eqnarray}}
\newcommand{\ee}{\end{eqnarray}}
\newcommand{\bea}{\begin{eqnarray}}
\newcommand{\eea}{\end{eqnarray}}

\newcommand{\ba}{\begin{array}}
\newcommand{\ea}{\end{array}}
\newcommand{\bd}{\begin{displaymath}}
\newcommand{\ed}{\end{displaymath}}
\newcommand{\beq}{\begin{equation}}
\newcommand{\eeq}{\end{equation}}

\def\it{ {\em i.e.,\ }}

\def\q2 {q^2}

\def\bt{\begin{table}}
\def\et{\end{table}}

\definecolor{shilamagenta}{rgb}{0.8, 0.0, 0.8}

\definecolor{shilagreen}{rgb}{0.0, 0.5, 0.0}
\definecolor{shilacyan}{rgb}{0.0, 0.58, 0.71}

\definecolor{midnightblue}{rgb}{0.1, 0.1, 0.44}

\usepackage{hyperref}

\hypersetup{citecolor=blue,        
    linkcolor=red,  
    filecolor=magenta,      
    urlcolor=shilacyan  
}

\usepackage{cleveref}
\usepackage{slashed}

\title{New bounds on Memory Burdened Primordial Black Holes from Big Bang Nucleosynthesis}

\author[a,1]{Arnab Chaudhuri,}
\author[a,b,c,d,2]{Kazunori Kohri}
\author[e,f,g,3]{and Valentin Thoss}

\affiliation[a]{Division of Science, National Astronomical Observatory of Japan, Mitaka, Tokyo 181-8588, Japan.}
\affiliation[b]{The Graduate University for Advanced Studies (SOKENDAI), Mitaka, Tokyo 181-8588, Japan.}
\affiliation[c]{Theory Center, IPNS, KEK, 1-1 Oho, Tsukuba, Ibaraki 305-0801, Japan.}
\affiliation[d]{Kavli IPMU (WPI), UTIAS, The University of Tokyo, Kashiwa, Chiba 277-8583, Japan.}
\affiliation[e]{Universitäts-Sternwarte, Ludwig-Maximilians-Universität München, Scheinerstr. 1, 81679 Munich, Germany.}
\affiliation[f]{Max-Planck Institute for Extraterrestrial Physics, Giessenbachstr. 1, 85748 Garching, Germany.}
\affiliation[g]{Excellence Cluster ORIGINS, Boltzmannstrasse 2, 85748 Garching, Germany.}

\emailAdd{arnab.chaudhuri@nao.ac.jp}
\emailAdd{kazunori.kohri@gmail.com}
\emailAdd{vthoss@mpe.mpg.de}

\abstract{Primordial black holes (PBHs) with masses below $10^9$ grams are typically assumed to have negligible cosmological impact due to their rapid evaporation via Hawking radiation. However, the ``memory burden'' effect which is a quantum suppression of PBH evaporation can dramatically alter their decay dynamics. In this work, we revisit early-Universe constraints on ultralight PBHs in this mass range, demonstrating that memory burden significantly alters previous constraints. We compute new cosmological bounds from BBN that strongly limit the presence of ultralight PBHs in the early Universe. We report that the PBHs in the mass range $10^0$--$10^2~\mathrm{g}$ for $k=2$ are unconstrained by observations.}

\begin{document}
\maketitle
\flushbottom

\section{Introduction}
\label{sec:intro}

Primordial Black Holes (PBHs), proposed over half a century ago~\cite{Carr:1974nx}, have re-emerged as compelling candidates in the study of dark matter (DM)~\cite{Carr:1974nx,Hawking:1971ei,Ivanov:1994pa,Carr:2009jm,Bartolo:2018evs,Carr:2020gox,Carr:2020xqk,Jedamzik:2020ypm,Jedamzik:2020omx,Green:2020jor,Villanueva-Domingo:2021spv,Carr:2021bzv} and gravitational wave sources~\cite{Baumann:2007zm,Espinosa:2018eve,Kohri:2018awv,Cai:2018dig,Wang:2019kaf,Domenech:2019quo,Ragavendra:2020sop,Inomata:2023zup,Franciolini:2023pbf,Firouzjahi:2023lzg,Maity:2024odg}. Unlike many DM scenarios, PBHs arise naturally from gravitational collapse without the need for new particle species.

However, PBHs are subject to strong constraints, particularly due to Hawking radiation~\cite{Hawking:1974rv,Hawking:1975vcx}, which implies that PBHs with masses $M < 10^{15}~\mathrm{g}$ would have evaporated by the present epoch. While such light PBHs cannot account for DM, they could have played roles in early-Universe phenomena such as reheating~\cite{He:2022wwy,RiajulHaque:2023cqe,He:2024wvt}, leptogenesis~\cite{Calabrese:2023key,Gunn:2024xaq}, gravitational wave production~\cite{Inomata:2019ivs,Sugiyama:2020roc,Inomata:2020lmk,Domenech:2020ssp,Papanikolaou:2020qtd,Domenech:2021wkk,Papanikolaou:2022chm,Bhaumik:2022pil,Bhaumik:2022zdd,Balaji:2024hpu,Wang:2022nml}, destabilizing the Higgs~\cite{Kohri:2017ybt,Hayashi:2020ocn}, and DM generation~\cite{Chaudhuri:2023aiv,Green:1999yh,Lennon:2017tqq,Cheek:2021odj,Bernal:2022oha}.

Surviving PBHs with $M \gtrsim 10^{15}~\mathrm{g}$ are also constrained by observations such as $\gamma$-ray backgrounds~\cite{Page:1976wx,MacGibbon:1991vc}, Big Bang nucleosynthesis (BBN)~\cite{Kohri:1999ex,Carr:2009jm,Clark:2016nst,Carr:2020gox}, and lensing or dynamical bounds~\cite{Poulin:2016anj,Niikura:2017zjd,Croon:2020ouk,Griest:2013aaa}. A detailed summary can be found in~\cite{Carr:2020gox}. For a monochromatic mass function, PBHs can comprise all of DM only in the mass window $10^{17}~\mathrm{g} \lesssim M \lesssim 10^{23}~\mathrm{g}$.

The lower mass cutoff stems from the inverse mass dependence of Hawking temperature, $T \sim 1/M$, which accelerates evaporation~\cite{Hawking:1974rv}. 
This process also injects energy into the early Universe, potentially disrupting $\gamma$-ray backgrounds, CMB anisotropies, and BBN~\cite{Auffinger:2022khh}. Accordingly, only a small abundance of light PBHs is allowed.
However, some studies suggest that certain quantum effects can delay or suppress evaporation. One such effect is the ``memory burden,'' first proposed in~\cite{Dvali:2018xpy}, which becomes relevant when a black hole has lost roughly half of its initial mass. Hawking’s semiclassical results, valid in the $G \to 0$, $M \to \infty$ limit with fixed Schwarzschild radius, neglect backreaction from emitted radiation. As evaporation proceeds, this backreaction grows significant, modifying the dynamics~\cite{Dvali:2020wft}.

The memory burden suppresses late-time evaporation and has been further developed in~\cite{Thoss:2024hsr,Barman:2024ufm,Barman:2024kfj,Basumatary:2024uwo,Chianese:2024rsn,Chianese:2025wrk,Athron:2024fcj,Calabrese:2025sfh,Dvali:2024hsb,Zantedeschi:2024ram,Datta:2023vbs,Bandyopadhyay:2025ast,Boccia:2025hpm,Montefalcone:2025akm,Tan:2025vxp}. It implies that an old black hole with mass $M$ is not equivalent to a young one with the same mass: the suppression mechanism delays the final burst of evaporation, altering the timing and energy injection into the plasma.

As shown in~\cite{Thoss:2024hsr,Alexandre:2024nuo}, this opens the possibility for PBHs with initial masses below $10^{10}~\mathrm{g}$ to evade standard BBN bounds. In this work, we revisit BBN constraints in the presence of memory burden for \textit{ultralight PBHs} which have fully evaporated by the present day. This regime has not been tightly constrained and offers a previously overlooked window for viable PBH cosmology.

We show that evaporation suppression significantly alters the constraints from the abundances of light elements. Our results reveal a parameter space where evaporated PBHs remain compatible with observations, motivating a revised framework for future studies of such scenarios.

This paper is organized as follows: in Section~\ref{comp} we review the components of memory-burdened PBH evolution; in Section~\ref{sec:modified}, we present the modified BBN bounds in the ultralight regime; and we conclude in Section~\ref{conc}.

\section{Components of Memory Burden}
\label{comp}
To understand the influence of memory burden on PBH dynamics, it is crucial to first outline the key elements governing PBH formation and evolution. The mass of a primordial black hole (PBH) resulting from gravitational collapse is closely tied to the horizon size at the moment of formation. This relationship is expressed as~\cite{Fujita:2014hha,Masina:2020xhk}:
\begin{equation}
M_0 = \frac{4}{3}\,\pi\,\gamma\,\left(\frac{1}{H\left(T_0\right)}\right)^3\,\rho_\text{rad}\left(T_0\right)\,,
\end{equation}
where \( H(T_0) \) is the Hubble rate during the radiation-dominated era at the time of PBH formation, and \( \gamma \) is an efficiency factor defining the fraction of the total mass within the Hubble radius that collapses into PBHs. $M_{\rm{0}}$ is the mass of the PBH at formation.

The radiation temperature at the point of PBH formation is given by:
\begin{equation}
    T_0 = \left(\frac{45\,\gamma^2}{16g_\star(T_0)}\right)^{1/4}\,\left(\frac{M_P}{M_0}\right)^{1/2}\,M_P\,,
\end{equation}
where \( g_\star(T_0) \) represents the relativistic degrees of freedom in the thermal bath, and \( M_P \simeq 1.2 \times 10^{19} \) GeV is the Planck mass. 

The time of PBH formation is determined by:
\begin{equation}
t_0 = \frac{M_0}{\gamma M_P^2}\,,
\end{equation}
assuming a standard radiation-dominated Universe with \( H(t) = 1/(2t) \). The initial abundance of PBHs can be described using the dimensionless parameter \(\beta\), defined as:
\begin{equation}
    \beta \equiv \frac{\rho_{\text{BH}}(t_0)}{\rho_R(t_0)}\,,
\end{equation}
where \(\rho_{\text{BH}}(t_0)\) is the energy density of the PBHs, and \(\rho_R(t_0)\) is the energy density of radiation at the time of formation. This parameter quantifies the fraction of the total energy density present in the form of PBHs at their formation, serving as an important measure of their initial contribution to the energy budget of the Universe.

The quantity \( f_{\rm PBH,0} \) denotes the fraction of DM composed of PBHs at the time of their formation, which is related to the initial PBH abundance \( \beta_{\rm PBH} \) by
\begin{equation}
f_{\rm PBH,0} = \frac{\beta_{\rm PBH}}{\Omega_{\rm DM}}\,,
\end{equation}
where \( \Omega_{\rm DM} \) is the density parameter of dark matter at the time of PBH formation. Note that $f_{\rm PBH,0}=1$ does not imply that the PBHs make up the entire dark matter at present day due to their evaporation after formation.

In our analysis, we adopt standard cosmological parameters, including a Hubble rate of \( h = 0.67 \), the relativistic degrees of freedom at formation \( g_{*} = 106.75 \), and a value of \( \gamma = 1 \). The parameter \( \gamma \) represents the ratio of the final black hole mass \( M \) to the mass within the Hubble volume \( M_H \) at the time of formation, otherwise known as the efficiency factor, as mentioned above. For more details, see Equations (2)–(6) in \cite{Carr:2020gox}.

The semiclassical emission rate of particles from a black hole is a cornerstone of Hawking radiation theory, describing how particle species \( i \) are emitted with energy \( E \) from a black hole of mass \( M \) and temperature \( T \) \cite{Hawking:1975vcx}. This rate is expressed as:  
\begin{equation}
\frac{\mathrm{d}^2 N_{i,\mathrm{SC}}}{\mathrm{d}E\mathrm{d}t}(E, M, s_i) = \frac{g_i}{2\pi} \frac{\Gamma(E, M, s_i)}{e^{E / T(M)} - (-1)^{s_i}}\,,
\label{eq:dndt_sc}
\end{equation}
where \( g_i \) represents the degrees of freedom of the particle species, \( s_i \) is the particle's spin, and \( \Gamma(E, M, s_i) \) are the greybody factors that quantify the deviation from a perfect black-body spectrum. The black hole's temperature \( T \) is intrinsically linked to its mass \( M \) through the relationship:  
\begin{equation}
    T = \frac{M_P^2}{8\pi M}\,.
\end{equation}  

The emission of particles leads to a gradual decrease in the black hole's mass, characterized by a mass loss rate:
\begin{equation}
\dot{M} = -\frac{\mathcal{F}(M)}{M^2}\,,
\end{equation}
where \( \mathcal{F}(M) \) is a function encapsulating the cumulative contributions of all particle species that the black hole emits at a given mass \( M \). The greybody factors and the function \( \mathcal{F}(M) \) are computed using the \textsc{BlackHawk} code \cite{Arbey:2019mbc}, which provides accurate predictions for the particle emission spectra of black holes across various scenarios.  
The total lifetime of a black hole within the Hawking framework is determined solely by its initial mass and is approximately given by:  
\begin{equation}
    t_{\rm SC}(M_0) \approx \frac{M_0^3}{3\mathcal{F}(M)}\,.
\end{equation}

In our revised framework for black hole evaporation, we incorporate modifications to Hawking's original model to account for deviations that arise during the later stages of evaporation. Specifically, we assume that Hawking's results hold up until the black hole reaches a critical mass \( M = qM_0 \), where \( q \) is the fraction of the initial black hole mass at which the standard semiclassical (SC) description begins to break down. Unless explicitly stated otherwise, we adopt \( q = 1/2 \), corresponding to the "half-decay" point.

Beyond this critical mass, the particle emission rate could be significantly suppressed due to a phenomenon called the "memory burden" (MB). This suppression is described in a parametrized form based on the insights of \cite{Dvali:2020wft}: 
\begin{equation}
	\frac{\mathrm{d}^2 N_{i,\mathrm{MB}}}{\mathrm{d}E \mathrm{d}t}(E, M_0, s_i) = \frac{1}{S(qM_0)^k} \frac{\mathrm{d}^2 N_{i,\mathrm{SC}}}{\mathrm{d}E \mathrm{d}t}(E, qM_0, s_i)\,,
	\label{eq:dndt_mb}
\end{equation}
where \( k \) is an adjustable parameter governing the degree of suppression, and \( S \) is the entropy of the black hole expressed in units of the Boltzmann constant \( k_B \). The entropy \( S \) is given by:
\begin{equation}
	S = \frac{4\pi M^2}{M_P} \approx 2.6 \times 10^{10} \left(\frac{M}{1\,\mathrm{g}}\right)^2\,.
\end{equation}

The extraordinarily large value of entropy for black holes implies that even modest values of \( k \) can lead to a suppression of the emission rate by several orders of magnitude. Importantly, under this modified framework, the black hole temperature and emission rate are assumed to remain constant in the MB regime. As a consequence, the decay of the black hole becomes linear, with the lifetime approximated as:
\begin{equation}
t_{\mathrm{MB}}(M) \approx \frac{(qM_0)^3 S(qM_0)^k}{\mathcal{F}(qM_0)}\,,
\label{eq:t_mb}
\end{equation}
where \( \mathcal{F}(qM_0) \) encapsulates the degrees of freedom of the emitted particles and \( S_0 \) is the initial entropy. This estimate neglects the contribution to the lifetime from the earlier SC phase, emphasizing the prolonged nature of black hole evaporation in the presence of memory burden effects.



\section{New bounds from BBN}
\label{sec:modified}
The primordial abundance of light elements serves as one of the earliest and most sensitive cosmological tools for investigating the effects of PBH evaporation. According to Hawking's theory, black holes with mass $M_0 \gtrsim 10^{10}$~g undergo evaporation during or after the epoch of Big Bang Nucleosynthesis (BBN)~\cite{Carr:2020gox}, the period when light elements such as hydrogen, helium, and lithium were formed. The energetic radiation and particles emitted during this process can significantly impact the delicate nuclear reactions driving the synthesis of these elements.

In particular, the radiation from evaporating PBHs can disrupt the neutron-to-proton ratio, a key parameter in determining the final abundances of helium and other light nuclei. Moreover, high-energy photons and hadrons emitted during evaporation can trigger photodissociation and hadrodissociation processes, fragmenting existing nuclei and altering the predicted elemental abundances.

The standard BBN framework, which assumes no additional energy injection, has been remarkably successful in matching the observed primordial abundances of light elements. As a result, any deviations caused by PBH evaporation are subject to stringent observational constraints. Modifications to the predicted abundances must remain minimal to avoid conflict with astrophysical and cosmological measurements, making the study of light element abundances a powerful probe for understanding PBH dynamics in the early Universe. According to the findings presented by \cite{Carr:2020gox}, which build upon earlier results from \cite{Carr:2009jm} and incorporate updates from \cite{Kawasaki:2004qu,Kawasaki:2017bqm} and \cite{Kawasaki:1999na,Kawasaki:2000en,Hasegawa:2019jsa}, the initial fraction of dark matter comprised of primordial black holes is tightly constrained to \( f_{\rm PBH,0} \sim 10^{-4} \) for $M_0\in[10^{10},10^{13}]$. The upper limit of this mass range is determined by the lifetime of the PBHs exceeding the radiation-dominated phase of the Universe. Beyond radiation-matter-equality, the processes responsible for the modification of the abundance of light elements become inefficient, leading to a weakening of the constraint.


The constraints imposed by BBN are primarily obtained by two particle emission modes: the radiative and the hadronic emission modes. The electromagnetic particles (photons, electrons, etc.) emitted from the evaporating PBH collide with background photons and electrons and form an electromagnetic shower. Ultimately, this shower thermalises and dissipates into the thermal bath. During the development of the electromagnetic shower, the low-energy particles inside the
electromagnetic shower dissociate light elements. The scale of the shower evolution is determined by the total amount of energy emitted from the evaporating PBH that ultimately thermalises. Therefore, the amount of light elements destroyed is determined by the total amount of energy emitted from the evaporating PBH. The dissociation of the light elements by the hadronic emission mode
occurs when the emitted particles are colored particles (quarks, gluons, etc.), which fragment to produce a lot of high-energy hadrons
(mesons and baryons). These trigger scattering with background particles (nuclei, nucleons, photons, electrons, etc.), forming a hadronic shower composed of energetic daughter hadrons. These daughter
hadrons eventually dissipate into the thermal bath through their decays and/or scatterings off the background particles. During the shower evolution, light elements are dissociated by those energetic
hadrons. The scale of this effect differs from that of the previous electromagnetic-shower case, as it is determined by the number of high-energy hadrons emitted just after the fragmentation. However, the
number of high-energy hadrons depends only very weakly on the energy of the initially emitted colored particles. According to the experimental data of the fragmentation to hadrons obtained by the
collider experiments, e.g., the Large Hadron Collider (LHC) experiment, this energy dependence does not increase linearly with energy but logarithmically.  More quantitatively, even when the
numbers of the high-energy hadrons emitted after the fragmentation are scaled by the total energy emitted from the PBH, the deviation is not significant.

For the reasons explained above, the physics of BBN is mostly sensitive to the total amount of energy injected by the PBHs at a given time after the Big Bang. Therefore, in order to compute bounds that take the memory burden into account we will assume that $f_{\rm PBH}$ is a function of the time of evaporation $t_{\rm ev}$ and the total energy injected only. We convert the existing bound $f_{\rm PBH,0}(M)$ to $f_{\rm PBH,0}(t_{\rm ev})$ according to their semiclassical lifetime. 
Then, we can obtain the bounds from the first phase of evaporation ($M_0\rightarrow qM_0)$ by rescaling the usual bounds by a factor of $1-q$ as this is the fraction of mass lost (and thus energy injected) before the memory burden sets in. For $q=0.5$, the full numerical calculation done by \cite{Thoss:2024hsr} found a relaxation of the bound by a factor of 0.55, close to the factor of 0.5 expected from the arguments above. The second stage of evaporation has a prolonged lifetime $t_{\rm MB}$ due to the memory burden effect, given by Equation~\ref{eq:t_mb}. We can compute the BBN bound for this phase according to $f'_{\rm PBH,0}(M)=qf_{\rm PBH,0}(t_{\rm ev}=t_{\rm MB})$, where $f'$ denotes the new bound. In summary, the constraint from BBN is computed by taking into account the lifetime of both evaporation stages and the fraction of the total energy released during these.

\begin{figure}[h]
  \centering
  \begin{minipage}[b]{0.48\textwidth}
    \includegraphics[width=\textwidth]{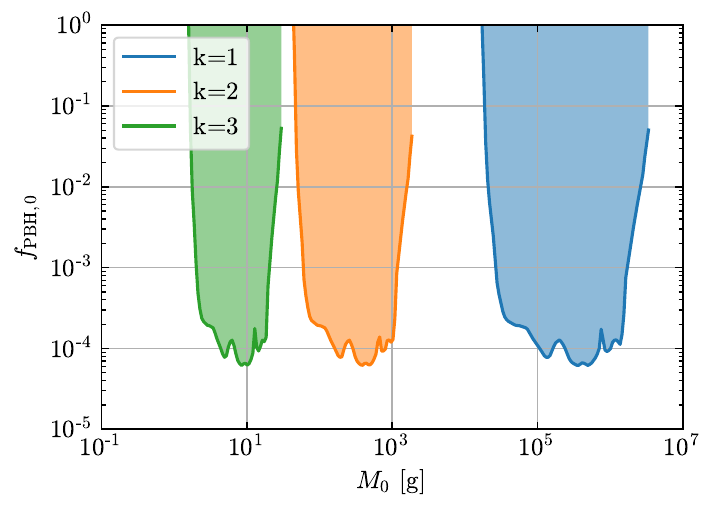} 
  \end{minipage}
  \hspace*{.1cm}
  \begin{minipage}[b]{0.48\textwidth}
    \includegraphics[width=\textwidth]{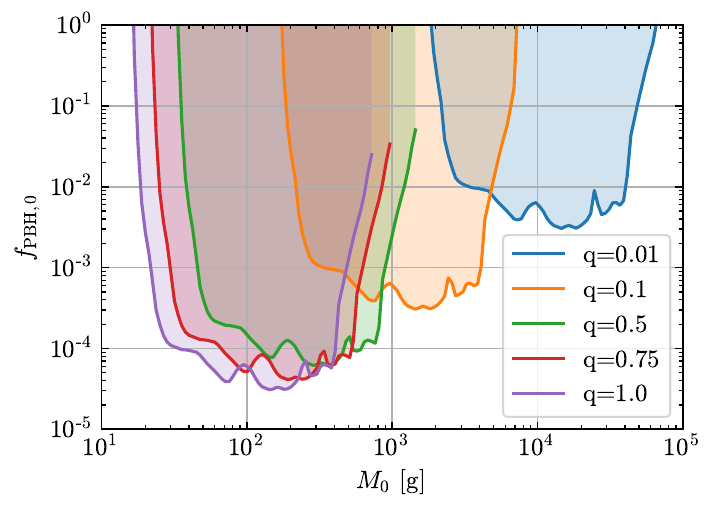}
      \end{minipage}
    \caption{Bounds on $f_{\rm{PBH,0}}(M_0)$ from BBN (shaded regions). The left panel displays constraints for $q=0.5$ and various values of $k$. The right panel shows bounds for $k=2$ and various value of $q$.}
 \label{bbn_1}
 \end{figure}

 \begin{figure}[h]
    \centering
    \includegraphics[width=\textwidth]{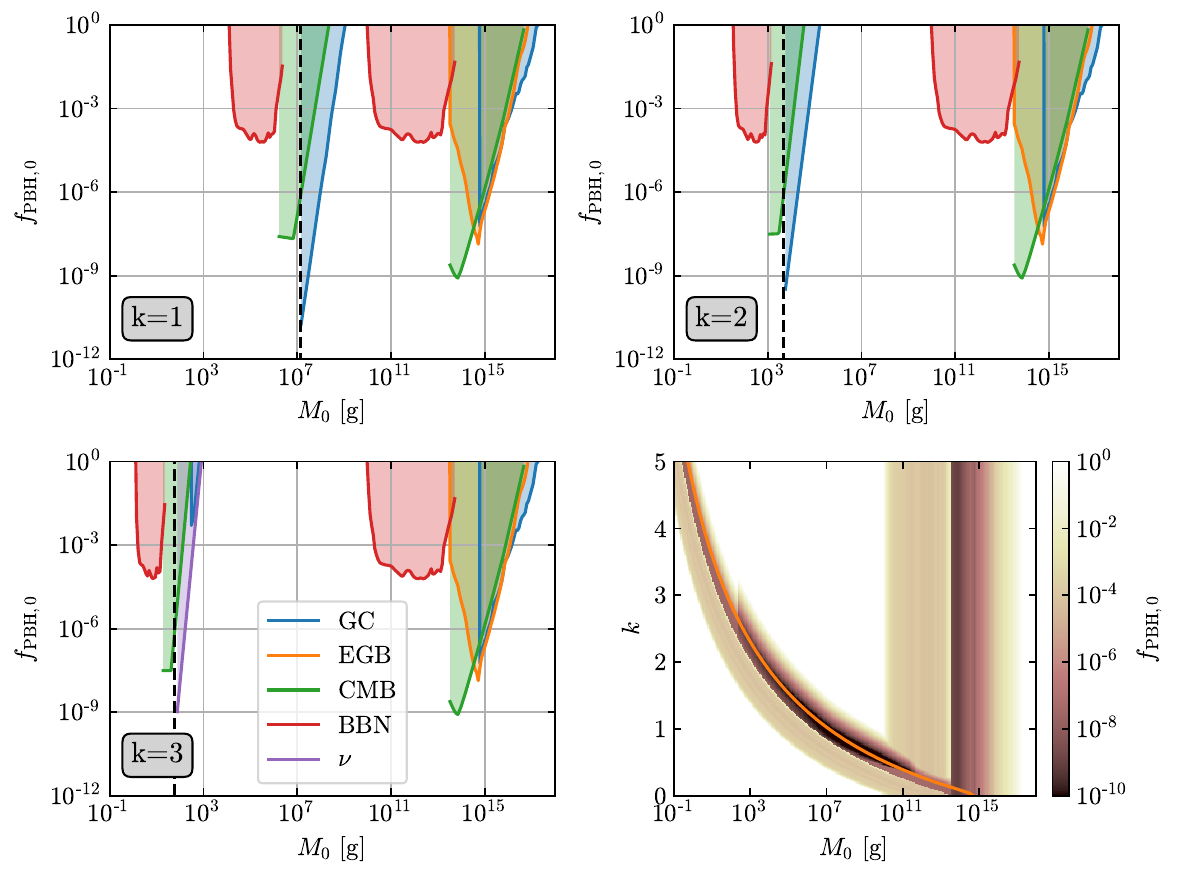}
    \caption{The top left / top right / bottom left panel show constraints on $f_{\rm{PBH,0}}(M_{0})$ for $k=1/2/3$. Displayed are bounds from $\gamma$ ray emission from the galactic center (GC), the extragalactic $\gamma$ ray background (EGB), CMB anisotropies and BBN. For the case of $k=3$ we also show the constraint from galactic neutrino emission from \cite{Chianese:2024rsn}. The vertical dashed line indicates the mass below which the PBHs fully evaporate by the present epoch. The bottom right panel shows the map of constraints on $f_{\rm{PBH,0}}$ as a function of $M_{0}$ and $k$ for $q=0.5$. For parameters below the orange line, the PBHs fully evaporate by the present day.}
    \label{bbn_2}
\end{figure}

In Figure~\ref{bbn_1} we display the bounds from BBN for various values of $q$ and $k$, that parametrize the memory burden effect. The prolongation of the evaporation means that the constraints now affect PBHs with much smaller initial masses, as they can now survive beyond the onset of BBN. Constraints from the semiclassical phase of evaporation restrict $f_{\rm PBH,0}<1$ for $M_{\rm PBH}\in[10^{10},10^{13}]\,\rm{g}$. Taking into account the memory burden, one obtains strong constraints for PBHs with $M_{\rm PBH}\in[10^{4},10^{6}]\,\rm{g}$ for $k=1$ and substantially lighter ones for stronger suppression, i.e. larger values of $k$. Notably, the width of the mass window constrained by BBN shrinks with increasing $k$ as a result of the stronger dependence of $t_{\rm MB}$ on $M_0$ (see Equation~\ref{eq:t_mb}).

Changing the value of $q$ shifts the bounds in mass according to $M_0\sim 1/q$ and affects the strength of the bound as $f_{\rm PBH,0}\sim q$. However, as the memory burden is expected to become relevant at latest when the black hole has lost half of its mass, values of $q\ll 0.5$ are physically of little interest. In the regime $q\in[0.5,1]$ the bounds are not overly sensitive to the exact value of $q$ and change at most by a factor of two.

Figure~\ref{bbn_2} compiles constraints from galactic and extragalactic $\gamma$ rays and CMB anisotropies, computed by \cite{Thoss:2024hsr} as well as the BBN bounds from this work for $k=1,2,3$. Note that we have removed the bound from extragalactic $\gamma$ rays for PBHs with mass $M_0<10^{10}\,\rm{g}$ as it was shown by \cite{Chianese:2025wrk} that attenuation is significant at these energies, which was neglected in previous work. In addition, as we do not trust the secondary emission rates for $E_{\rm sec}/T\ll 10^{-6}$, the constraints from $\gamma$ ray emission are limited to $M_0\gtrsim10^3\,\rm{g}$ for the observational data considered. However, there are comparable constraints for $M_0<10^{3}\,\rm{g}$ from neutrino experiments and ultra-high $\gamma$ ray observatories which have been computed by \cite{Chianese:2024rsn,Chianese:2025wrk}. Therefore, we have added the bound from neutrino emission of \cite{Chianese:2024rsn} for the case of $k=3$. For lower values of $k$ the resulting bound is very similar to the one from the galactic $\gamma$ ray emission and we omit it for the sake of readability.

The bounds from BBN extend the existing constraints to lighter PBHs that do not survive to the present day - unless they leave behind relics. While they are thus excluded from making up the present dark matter, light evaporating PBHs have been studied as a mechanism to produce particle DM, to address baryogenesis or as a source of gravitational waves \cite{Hawking:1975vcx, Dolgov:2000ht, Baumann:2007zm, Fujita:2014iaa, Allahverdi:2017sks, Lennon:2017tqq,Keith:2020jww,Coogan:2019qpu,Baldes:2020nuv, Acharya:2020jbv,Masina:2021zpu,Papanikolaou:2020qtd,Papanikolaou:2023cku}, recently also in the context of the memory burden effect \cite{Bhaumik:2024qzd,Barman:2024iht,Haque:2024eyh,Balaji:2024hpu,Kohri:2024qpd,Barman:2024ufm,Barman:2024kfj,Athron:2024fcj,Loc:2024qbz,Borah:2024bcr,Barker:2024mpz,Jiang:2024aju,Calabrese:2025sfh}. For these analyses, our constraints provide important limits for the available parameter space. In particular, for $k=1$, PBHs have to be lighter than $M_0=10^4\,\rm{g}$ in order to avoid cosmological constraints and evaporate before the onset of BBN. Observational limits on the tensor-to-scalar ratio imply that PBHs, which form from density fluctuations seeded by inflation, have a mass of at least $M_0\sim 1\,\rm{g}$ \cite{Planck:2018jri}. Together with our results, this imposes a bound of $k\lesssim 3$ in order to have fully evaporating PBHs that are not strongly constrained by BBN, although the precise bound will depend on the value of $\gamma$ and the accretion of the black hole after horizon formation.

\section{Conclusion}
\label{conc}
The suppression of Hawking evaporation due to the memory burden effect fundamentally reshapes constraints on PBHs, most notably for $M_0<10^{10}\,\rm{g}$ - assuming it becomes relevant when the black hole has lost half its initial mass. Previous work has focused on the question, whether this opens up a new window for PBHs to make up the entire dark matter. However, even much lighter PBHs that cannot explain the present dark matter density can be interesting probes of the physics of the early Universe.

In this work we compute the bounds on ultralight PBHs arising from their effect on BBN. This early-Universe probe extends previous constraints towards lower PBH masses and poses strong limits on their parameter space.

These results highlight the importance of incorporating the memory burden effects into the modelling of evaporating black holes in the early Universe. This has implications for PBH phenomenology and provides a novel setting in which to test predictions of quantum gravity in a cosmological context.

Future constraints from improved measurements of primordial deuterium and helium abundances, as well as searches for late-time energy injection via spectral distortions, gamma-ray backgrounds, or CMB anisotropies, will be critical in further probing the existence of light PBHs. A more complete theoretical understanding of the memory burden mechanism and its parameter dependence will also be essential to delineate the precise range of allowed PBH masses. Our results motivate continued exploration of quantum-corrected black hole evaporation and its implications for early-Universe physics.

\section{Acknowledgment}
The work of Arnab Chaudhuri was supported by the Japan Society for the Promotion of Science (JSPS) as a part of the JSPS Postdoctoral Program (Standard), grant number JP23KF0289. The work of Valentin Thoss was supported by the Excellence Cluster ORIGINS which is funded by the Deutsche Forschungsgemeinschaft (DFG, German Research Foundation) under Germany’s Excellence Strategy - EXC-2094 - 390783311. The work of Kazunori Kohri was partly supported by KAKENHI Grant Nos. JP24H01825, and JP24K07027.

\end{document}